\documentstyle[aps,prl,preprint,floats,epsfig]{revtex}

\newcommand{\raw}{\rightarrow}

\newcommand{\beq}{\begin{equation}}
\newcommand{\eeq}{\end{equation}}
\newcommand{\bqa}{\begin{eqnarray}}
\newcommand{\eqa}{\end{eqnarray}}
\newcommand{\bea}{\begin{array}}
\newcommand{\ena}{\end{array}}

\newcommand{\pip}{\mbox{$\pi^+$}}
\newcommand{\pim}{\mbox{$\pi^-$}}
\newcommand{\pipm}{\mbox{$\pi^\pm$}}
\newcommand{\pimp}{\mbox{$\pi^\mp$}}
\newcommand{\piz}{\mbox{$\pi^0$}}

\newcommand{\rhop}{\mbox{$\rho^+$}}
\newcommand{\rhom}{\mbox{$\rho^-$}}

\newcommand{\rhomp}{\mbox{$\rho^{\mp}$}}
\newcommand{\rhoz}{\mbox{$\rho^0$}}

\newcommand{\etpri}{\mbox{$\eta'$}}
\newcommand{\omg}{\mbox{$\omega$}}

\newcommand{\kzs}{\mbox{$K^{0}_{S}$}}
\newcommand{\kb}{\mbox{$\bar{K^0}$}}

\newcommand{\kp}{\mbox{$K^+$}}
\newcommand{\km}{\mbox{$K^-$}}
\newcommand{\kpm}{\mbox{$K^\pm$}}

\newcommand{\kstm}{\mbox{$K^{*-}$}}

\newcommand{\kstz}{\mbox{$K^{*0}$}}
\newcommand{\kstb}{\mbox{$\bar{K}^{*0}$}}

\newcommand{\dz}{\mbox{$D^0$}}

\newcommand{\bp}{\mbox{$B^+$}}
\newcommand{\bm}{\mbox{$B^-$}}

\newcommand{\bz}{\mbox{$B^0$}}
\newcommand{\bzb}{\mbox{$\bar{B}^0$}}

\begin{document}

\tighten

\begin{flushright}
{CLNS 99/1652} \\
{CLEO 99-19} \\

\end{flushright}

\vskip 0.5cm

\begin{center}

{\large \bf Study of Charmless Hadronic $B$ Meson Decays}

\vskip 0.3cm

{\large \bf  to  Pseudoscalar-Vector Final States}

\end{center}

\vskip 1.0cm

\begin{center}
{\large  CLEO Collaboration}
\end{center}

\vskip 0.4cm

\begin{center}
{\large  (August 4, 2000)}
\end{center}

\vskip 1.5cm

\begin{center}
{\Large  Abstract}
\end{center}

\vskip 0.3cm

We report results of searches for charmless hadronic $B$ meson decays to
pseudoscalar($\pipm$, $\kpm$, $\piz$ or $\kzs$)-vector($\rho$, $K^*$ or $\omg$)
final states.
Using $9.7 \times 10^6$ $\mbox{$B\bar{B}$}$ 
pairs collected with the CLEO detector, 
we report first observation of
$\bm\raw\pim\rhoz$, $\bzb\raw\pipm\rhomp$ and $\bm\raw\pim\omg$, 
which are expected to be dominated by hadronic $b$ $\raw$ $u$ transitions.
The measured branching fractions are 
(10.4$^{+3.3}_{-3.4} \pm 2.1) \times 10^{-6}$,
(27.6$^{+8.4}_{-7.4} \pm 4.2) \times 10^{-6}$ and
(11.3$^{+3.3}_{-2.9} \pm 1.4) \times 10^{-6}$, respectively.
Branching fraction upper limits are set for all the other decay
modes investigated.

\vskip 0.5cm

PACS numbers: 13.20.He,13.25.-k,13.25.Hw,13.30.Eg,14.40.Nd

\newpage
\begin{center}
C.~P.~Jessop,$^{1}$ H.~Marsiske,$^{1}$ M.~L.~Perl,$^{1}$
V.~Savinov,$^{1}$ D.~Ugolini,$^{1}$ X.~Zhou,$^{1}$
T.~E.~Coan,$^{2}$ V.~Fadeyev,$^{2}$ Y.~Maravin,$^{2}$
I.~Narsky,$^{2}$
% ...
\end{center}

\small
\begin{center}
$^{1}${Stanford Linear Accelerator Center, Stanford University, Stanford,
California 94309}\\
$^{2}${Southern Methodist University, Dallas, Texas 75275}\\
$^{3}${Syracuse University, Syracuse, New York 13244}\\
$^{4}${University of Texas, Austin, TX  78712}\\
% ...
$^{6}${Virginia Polytechnic Institute and State University,
Blacksburg, Virginia 24061}\\
$^{7}${Wayne State University, Detroit, Michigan 48202}\\
$^{8}${California Institute of Technology, Pasadena, California 91125}\\
$^{9}${University of California, San Diego, La Jolla, California 92093}\\
$^{10}${University of California, Santa Barbara, California 93106}\\
% ...
\end{center}

\newpage

$CP$ violation in the Standard Model (SM) is a consequence of
the complex phase in the Cabibbo-Kobayashi-Maskawa (CKM) 
quark-mixing matrix~\cite{ckm}. 
The study of charmless hadronic decays of $B$ mesons plays a key role
in testing the SM picture of $CP$ violation. 
For example, the angle 
$\alpha$ $\equiv$ $arg$ [($- V_{td}V_{tb}^{*})/(V_{ud}V_{ub}^{*})$] 
of the unitarity triangle can be measured
by performing a full Dalitz analysis of the decays
$\bz(\bzb)\raw\pip\rhom$, $\pim\rhop$ and $\piz\rhoz$~\cite{helen_prd93}.
While the CLEO data do not yet have the sensitivity for the $CP$ violation
measurements, experimental measurements of these decay modes
will be useful to test various theoretical predictions that typically 
make use of effective Hamiltonians, often  with factorization
assumptions~\cite{sum}.
Recently, it has been suggested~\cite{largegamma}, with model dependency, that
published experimental results on charmless hadronic $B$ decays 
indicate that $\cos{\gamma} < 0$, in disagreement with current fits to 
the information most sensitive to CKM matrix elements~\cite{ckmfits}.

In this Letter, we present results of searches for
$B$ meson decays to exclusive pseudoscalar-vector ($B \raw PV$) final states 
that include a pseudoscalar meson $\pipm$, $\kpm$, $\piz$ or $\kzs$ and
a vector meson $\rho$, $K^*$ or $\omg$.
In particular we present first observation of the decays
$\bm\raw\pim\rhoz$, $\bzb\raw\pipm\rhomp$ and $\bm\raw\pim\omg$
(charge-conjugate modes are implied) which are expected to be 
dominated by hadronic $b$ $\raw$ $u$ transitions.
Our results supersede previous CLEO results on these decay 
modes~\cite{omega,bigrare}.

The data were collected with two configurations 
(CLEO II~\cite{CLEOII} and CLEO II.V~\cite{CLEOII.V}) of the
CLEO detector at the Cornell Electron Storage Ring (CESR). 
They consist of 9.1 fb$^{-1}$ taken at the $\Upsilon$(4S), which
corresponds to $9.7 \times 10^6$ $\mbox{$B\bar{B}$}$ pairs, 
and 4.4 fb$^{-1}$ taken below $B\overline{B}$ threshold, used for continuum
background studies. The data sample contains a factor of 3 more statistics 
than previously published results~\cite{omega}. 
In addition, the CLEO II data were reanalyzed with improved calibration and 
track-fitting, allowing for larger geometric acceptance and more efficient
track quality requirements.

The final states of the decays under study are reconstructed by
combining detected photons and charged pions and kaons. 
The detector elements most important for the results presented 
here are the tracking system, which consists of several  
concentric detectors operating inside a 1.5 T superconducting solenoid, 
and the high-resolution electromagnetic calorimeter, consisting of 7800 CsI(Tl)
crystals. For CLEO II, the tracking system consists of a 6-layer
straw tube chamber, a 10-layer precision drift chamber, and a
51-layer main drift chamber. The main drift chamber also provides a
measurement of the specific ionization loss, $dE/dx$, used for
particle identification.  For CLEO II.V the straw tube
chamber was replaced by a 3-layer, double-sided silicon vertex
detector, and the gas in the main drift chamber was changed from 
an argon-ethane to a helium-propane mixture.

 The resonances in the final state are identified via the decay modes 
 $\rho\raw\pi\pi$, $K^{*} \raw K\pi$ ($K^{*0}\raw\kp\pim$,
 $K^{*+}\raw\kp\piz$) and $\omg\raw\pip\pim\piz$.
 Reconstructed charged tracks are required to pass quality cuts based
 on their track fit residuals and impact parameter in both the
 $r$--$\phi$ and $r$--$z$ planes, and on the number of main drift chamber
 measurements. Each event must have a total of at least four such
 charged tracks. The $dE/dx$ measured by the main drift chamber
 is used to distinguish kaons from pions. 
 Electrons are rejected based on $dE/dx$ information and the 
 ratio of the measured 
 track momentum and the associated shower energy in the calorimeter. 
 Muons are rejected by requiring that charged tracks penetrate 
 fewer than seven interaction lengths of material. 
 Pairs of charged tracks used to reconstruct $\kzs$ 
 (via $\kzs \rightarrow \pi^+ \pi^-$) are required to have a common 
 vertex displaced from the primary interaction point. The invariant 
 mass of the two charged pions is required to be within two standard 
 deviations ($\sigma$) of the known $\kzs$ mass~\cite{pdg}. 
 Furthermore, the $\kzs$ momentum vector, obtained from
 a kinematic fit of the charged pions' momenta, is
 required to point back to the beam spot.
 To form $\piz$ candidates, pairs of photon candidates with an invariant mass 
 within 2.5$\sigma$ of the nominal $\piz$ mass are kinematically fitted 
 with the mass constrained to the known $\piz$ mass~\cite{pdg}.

 The primary means of identification of $B$ meson candidates is
 through their measured mass and energy. The beam-constrained mass of 
 the candidate is defined as $M_{B} \equiv \sqrt{E_b^2 - |{\bf p}|^2}$, 
 where $\bf p$ is the measured momentum of the candidate and $E_{b}$ is
 the beam energy. The resolution of $M_{B}$ ranges from 2.5 to 3.5 MeV, 
 where the larger resolutions correspond to decay modes with neutral pion(s). 
 The second observable $\Delta E$ is defined as 
 $\Delta E \equiv E_1 + E_2 - E_b$, 
 where $E_1$ and $E_2$ are the energies of the two final state mesons.
 The resolution of $\Delta E$ is mode dependent. For final states without a
 neutral pion, the $\Delta E$ resolution is about 20 MeV. For decay modes
 with one or two energetic neutral pions ($\bzb\raw\pipm\rhomp$,
 $\bzb\raw\piz\rhoz$ and $\bm\raw\piz\rhom$ etc), the $\Delta E$ 
 resolution worsens by
 approximately a factor of 2 or 3 and becomes slightly asymmetric 
 because of energy loss out of the back of the CsI crystals.
 We accept events with $M_{B}$ $>$ 5.2 GeV and $|\Delta E |$
 $<$ 100 to 300 MeV depending on the decay mode. 

 The vector meson $\rho$, $K^*$ and $\omg$ candidates are 
 required to have masses within 200, 75 and 50 MeV of their 
 known masses~\cite{pdg}, respectively. 
 In the simultaneous analysis of $\bzb\raw\piz\rhoz$ and $\piz\kstz$, 
 the $\rhoz$ or $\kstz$ candidate is required to have mass between 
 0.3 GeV to 1.0 GeV under the $\pip\pim$ decay hypothesis
 so that both $\rhoz$ and $\kstz$ enter into the sample.
 Because of the polarization of the vector meson, the soft decay product 
 from the vector meson may have momentum as low as 150 MeV. 
 To reduce the large combinatoric background from soft $\piz$s, only half 
 of the helicity (${\cal H}$) range, 
 corresponding to a hard $\piz$, is selected when a 
 $\rhop$ or $K^{*+}$ decays to a $\pip\piz$ or $\kp\piz$.
 The helicity is defined as the cosine of the angle between one of the 
 vector meson decay products in the vector meson rest frame and the 
 direction of the vector meson momentum in the lab frame.

 The main background comes from continuum $e^{+}e^{-}$ $\raw$ 
 $q\bar q$, where $q$ $=$ $u$, $d$, $s$, $c$. 
 This background typically exhibits a two-jet structure and
 can be reduced with event shape criteria.  
 We calculate the angle $\theta_{S}$ ($\theta_{T}$) between the sphericity 
 axis~\cite{sphericity} (thrust axis~\cite{thrust}) of the candidate 
 and the sphericity axis (thrust axis) of the rest of the event.
 The distribution of $\cos\theta_{S}(\theta_{T})$ should be flat for 
 $B$ mesons and strongly peaked at $\pm$1.0 for continuum background. 
 We require $| \cos\theta_{S}|$ $<$ 0.8 when there is a $\rho$ or $K^*$ 
 meson in the final state, and $| \cos\theta_{T}|$ $<$ 0.8 when there
 is a $\omg$ meson in the final state. 
 We also form a Fisher discriminant (${\cal F}$) with event shape
 observables~\cite{bigrare}.

 We then perform unbinned maximum-likelihood fits where the likelihood 
 of an  event is parameterized by the sum of probabilities for
 all relevant signal and background hypotheses, with relative weights 
 determined by maximizing the likelihood function ($\cal L$)
 \cite{omega,bigrare}.
 The probability of a particular hypothesis is calculated as a product 
 of the probability density functions (PDFs) for each of the input 
 observables.
 The observables used in the fit are $\Delta E$, $M_{B}$, ${\cal F}$, 
 ${\cal H}$ and the invariant mass of the resonance candidate. 
 For final states with the same vector meson but different charged
 light mesons (pion or kaon), we also use the 
 $dE/dx$ measurement of the high-momentum track and fit for both 
 modes simultaneously. 
 Similarly, 
 $dE/dx$ measurements of the vector meson decay daughters are used 
 in the simultaneous fit for $\bzb\raw\piz\rhoz$ and $\piz\kstz$.
 For each decay mode investigated, the signal PDFs are determined 
 with fits to GEANT-based simulation~\cite{geant} samples.
 The parameters of the continuum background PDFs are 
 determined with similar fits to simulated continuum samples as well as 
 continuum data. 
 Simulated continuum distributions are in excellent agreement with 
 the data taken below the $B\overline{B}$ threshold.
 Correlations between observables used in the fits
 are investigated and their effect is found to be negligible.

 In all cases, the fit includes hypotheses for 
 signal decay modes and the dominant continuum background. 
 Using the PDFs formed by the above observables, signal and 
 continuum background can be well separated. 
 For a few channels where the selected sample contains contributions
 from other $B$ decays, we also include hypotheses for background 
 from other $B$ decay modes. 
 These background decay modes can also be separated efficiently
 from the signal decay modes. 
 We select a sample that contains both $\bm\raw\pim\rhoz$, $\km\rhoz$ 
 and some contamination from $\bm\raw\pim\kstb$.
 We then fit simultaneously for $\bm\raw\pim\rhoz$, $\km\rhoz$ 
 with and without a $\bm\raw\pim\kstb$ contribution.
 Similarly, we select a sample that contains both $\bm\raw\pim\kstb$,
 $\km\kstb$ with some contamination from $\bm\raw\pim\rhoz$,
 $\km\rhoz$. Then we perform a simultaneous fit for
 $\bm\raw\pim\kstb$, $\km\kstb$ with or without the 
 $\bm\raw\pim\rhoz$, $\km\rhoz$ contributions.
 In both cases the fits with and without the background modes are 
 consistent with each other.
 For each of the combinations $\bzb\raw\piz\rhoz$, $\piz\kstz$, 
 $\bzb\raw\pipm\rhomp$, $\kpm\rhomp$, and
 $\bm\raw\pim\omg$, $\km\omg$, contributions from other $B$ decays
 are negligible and we select a common sample to fit for both modes.
 Finally individual samples are selected and fit for the $\bm\raw\piz\rhom$,
 $\bm\raw\piz\kstm$, $\bzb\raw\piz\omg$ and $\bzb\raw\kzs\omg$ searches.

 The contributions of $b$ $\raw$ $c$ and other $B$ decays are small
 in the selected samples of final states containing three tracks or two 
 tracks and a $\piz$, and their effects on the signal yields are negligible, 
 except in the samples of $\bm\raw\pim\rhoz$, $\km\rhoz$
 and $\bm\raw\pim\kstb$, $\km\kstb$.
 Events from $\bm\raw\dz\pim$ where 
 $\dz\raw\kpm\pimp, \pip\pim$ can enter 
 into these samples and mimic our signal. 
 We therefore impose a 30 MeV ($\sim$ 4$\sigma$) wide 
 $\dz\raw\pip\pim, \kpm\pimp$ invariant mass veto in all the 
 charged track pair combinations.
 We have also studied background from $\bm\raw\km\etpri$, with 
 $\etpri\raw\rhoz\gamma$~\cite{etapri,pdg}. 
 This background has exactly the same final state
 particles as $\bm\raw\km\rhoz$ with an extra photon.
 Approximately 3\% of this background can pass the selection for the
 $\bm\raw\pim\rhoz$, $\km\rhoz$ sample, therefore we include a component 
 in the fit to describe this contribution.
 For $\bm\raw\piz\rhom$ and $\bm\raw\piz\kstm$ modes, 
 due to the limited $\Delta E$ resolution for the final
 state with two neutral pions, the selected sample may
 contain background from other $B$ processes such as
 $B\raw\pi a_{1}$, $\rho\rho$.

 Table~\ref{results1} shows the results of these measurements. 
 The one standard deviation statistical 
 error on the yield is determined by finding the ranges for
 which the quantity $\chi^2 = -2\ln{\cal L}$ changes 
 by one unit.
 We observe significant yields for the decays $\bm\raw\pim\rhoz$, 
 $\bzb\raw\pipm\rhomp$, $\bm\raw\pim\omg$ and $\bm\raw\piz\rhom$.
 To verify that the yields we observe in $B$ meson decays to three-pion 
 final states are indeed due to $\pi\rho$ decays, we repeat 
 the standard fit allowing for an additional three-pion
 ``non-resonant" contribution. The PDFs for this contribution are 
 identical to the ones used for $B\raw\pi\rho$ signals except that 
 we use PDFs that are constants in the $\rho$ mass and ${\cal H}$. We find 
 that this has no effect on the yield and the significance for 
 $\bm\raw\pim\rhoz$ 
 and $\bzb\raw\pipm\rhomp$ signals. 
 Possible contributions from all other $B$ processes, including higher mass 
 pseudoscalar-vector decays, were also investigated for these channels 
 and found to be negligible.  
 However, the signal yield for  
 $\bm\raw\piz\rhom$ drops from $23.7^{+8.4}_{-7.4}$ with a significance
 of 5.1$\sigma$ to $8.0^{+9.1}_{-7.9}$ events with a 
 significance of only 1$\sigma$. 
 We can not rule out the possibility that a significant fraction
 of the observed yield in $\piz\rhom$ comes from 
 poorly measured processes such as non-resonant $\pim\piz\piz$, 
 $\pi a_{1}$ and $\rho\rho$ processes~\cite{pdg}.
 Therefore we calculate a conservative upper limit on the branching
 fraction assuming that the observed yield is due to  
 $\bm\raw\piz\rhom$ decays only.

\begin{table}[hhh]
\caption{Measurement results. Displayed are the decay mode, 
 event yield from the fit, total efficiency including secondary
 branching fraction $\epsilon$, statistical significance ($\sigma$),
 branching fraction from the fit ${\cal B}_{fit}$ (in units of 10$^{-6}$), 
 the measured branching fraction (${\cal B}$) or 90\% confidence 
 level upper limit (in units of 10$^{-6}$) and theoretical 
 prediction~\protect\cite{sum} (in units of 10$^{-6}$). 
 For the branching fraction measurement, the first error is
 statistical and the second systematic.
 We assume equal branching fractions for $\Upsilon (4S) \raw \bz\bzb$
 and $\bp\bm$.}
\begin{center}
\begin{tabular}{ l c c c c c c } 
Decay mode & Yield & $\epsilon$(\%) & Signif. & 
${\cal B}_{fit}$ & ${\cal B}$ or 90\% ${\cal B}$ UL & Theory        \\ 
$\bm\raw\pim\rhoz$         & $29.8^{+9.3}_{-9.6}$   & 30  & 5.4 &
                             $10.4^{+3.3}_{-3.4} \pm 2.1$       &
                             $10.4^{+3.3}_{-3.4} \pm 2.1$       &
                             0.4 $-$ 13.0                           \\ 
$\bm\raw\km\rhoz$          & $22.4^{+10.7}_{-9.1}$  & 28  & 3.7 &
                             $8.4^{+4.0}_{-3.4} \pm 1.8$        &
                             $<$ 17                             &
                             0.0 $-$ 6.1                            \\ 
$\bm\raw\pim\kstb$         & $13.4^{+6.2}_{-5.2}$   & 18  & 3.6 &
                             $7.6^{+3.5}_{-3.0} \pm 1.6$        &         
                             $<$ 16                             &
                             3.4 $-$ 13.0                           \\ 
$\bm\raw\km\kstb$          & $0.0^{+2.2}_{-0.0}$    & 17  & 0.0 &
                             $0.0^{+1.3+0.6}_{-0.0-0.0}$ &
                             $<$ 5.3                            &
                             0.2 $-$ 1.0                            \\ 
$\bzb\raw\pipm\rhomp$      & $31.0^{+9.4}_{-8.3}$   & 12  & 5.6 &
                             $27.6^{+8.4}_{-7.4} \pm 4.2$       &
                             $27.6^{+8.4}_{-7.4} \pm 4.2$       &
                             12 $-$ 93                              \\ 
$\bzb\raw\kpm\rhomp$       & $16.4^{+7.8}_{-6.6}$   & 11  & 3.5 &
                             $16.0^{+7.6}_{-6.4} \pm 2.8$       &
                             $<$ 32                             &
                             0.0 $-$ 12.0                           \\  
$\bzb\raw\piz\rhoz$        & $5.4^{+6.5}_{-4.8}$    & 34  & 1.2 &
                             $1.6^{+2.0}_{-1.4} \pm 0.8$        &
                             $<$ 5.5                            &
                             0.0 $-$ 2.5                            \\ 
$\bzb\raw\piz\kstb$        & $0.0^{+3.0}_{-0.0}$    & 25  & 0.0 &
                             $0.0^{+1.3+0.5}_{-0.0-0.0}$ &
                             $<$ 3.6                            &
                             0.7 $-$ 6.1                            \\  
$\bm\raw\piz\rhom$         & $23.7^{+8.4}_{-7.4}$   & 10  & 5.1 &
                             See text                           &
                             $<$ 43                             &
                             3.0 $-$ 27.0                           \\ 
$\bm\raw\piz\kstm$         & $2.6^{+4.2}_{-2.6}$    & 4   & 1.0 &
                             $7.1^{+11.4}_{-7.1} \pm 1.0$       &
                             $<$ 31                             &
                             0.5 $-$ 24.0                           \\ 
$\bm\raw\pim\omg$          & $28.5^{+8.2}_{-7.3}$   & 26  & 6.2 &
                             $11.3^{+3.3}_{-2.9} \pm 1.4$       &
                             $11.3^{+3.3}_{-2.9} \pm 1.4$       &
                             0.6 $-$ 24.0                           \\ 
$\bm\raw\km\omg$           & $7.9^{+6.0}_{-4.7}$    & 26  & 2.1 &
                             $3.2^{+2.4}_{-1.9} \pm 0.8$        &
                             $<$ 7.9                            &
                             0.2 $-$ 14.0                           \\ 
$\bzb\raw\piz\omg$         & $1.5^{+3.5}_{-1.5}$    & 19  & 0.6 &
                             $0.8^{+1.9+1.0}_{-0.8-0.8}$  &
                             $<$ 5.5                            &
                             0.0 $-$ 12.0                           \\ 
$\bzb\raw\kb\omg$          & $7.0^{+3.8}_{-2.9}$    &  7  & 3.9 &
                             $10.0^{+5.4}_{-4.2} \pm 1.4$       &
                             $<$ 21                             &
                             0.0 $-$ 17.0                           \\ 
\end{tabular}
\end{center}
\label{results1}
\end{table}

 Fig. 1 shows the likelihood contours from fits to  
 $\bm\raw\pim\rhoz$, $\km\rhoz$, $\bzb\raw\pipm\rhomp$, $\kpm\rhomp$ 
 and $\bm\raw\pim\omg$, $\km\omg$. The resulting branching fractions
 are given in Table I.
 Fig. 2 shows the $M_{B}$ and $\Delta E$ distributions after further 
 requirements are made on event probability to reduce background. 
 For the remaining processes in Table I we do not consider the signal 
 yields to be significant (i.e. significance drops to less than 3 
 $\sigma$ after all the possible systematics are taken into account), and 
 therefore set 90\% C. L. upper limits for their branching fractions.
 Note that for the $\bm\raw\km\omg$ decay mode the additional CLEO II.V
 data and the re-analysis of CLEO II data no longer support its previously 
 reported observation~\cite{omega}. However, the combined branching fraction
 ${\cal B}(\bm\raw h^{-}\omg$) = $(14.3^{+3.6}_{-3.2} \pm 2.0)
 \times 10^{-6}$ (where $h$ = $K$ or $\pi$) is still consistent with 
 the previous result.

 Systematic errors are separated into two categories. The first
 consists of systematic errors in the PDFs, which are determined by
 varying the PDF parameters within their uncertainty.
 The second consists of systematic errors associated with 
 event selection and efficiency factors. These are determined with 
 studies of independent data samples.
 For branching fraction central values, the systematic error is the 
 quadrature sum of the two components. 
 For upper limits, the likelihood function is integrated
 to find the yield value that corresponds to 90\% of the total
 area. The PDF systematic errors are taken into account
 in this procedure. The selection efficiency is then reduced by one standard 
 deviation when calculating the final upper limit.
 As a goodness-of-fit check we compare $-2\ln{\cal L}$ at the minimum 
 for our fits with expectations from fits to Monte Carlo experiments,
 and find them to be consistent in all cases.

 In summary, we have made first observation of the decays
 $\bm\raw\pim\rhoz$, $\bzb\raw\pipm\rhomp$ and $\bm\raw\pim\omg$.
 All of these $\Delta S= 0$ decay modes are expected to be dominated by
 hadronic $b$ $\raw$ $u$ transitions. We see no significant yields
 in any of the $\Delta S =1$ transitions. This is in contrast to
 the corresponding charmless hadronic $B$ decays to two pseudo-scalar 
 mesons ($B \raw PP$) $B\to K\pi$, $\pi\pi$, where $\Delta S = 1$ 
 transitions clearly dominate~\cite{kpi-pipi-paper}.
 It indicates that gluonic penguin decays play less of a role in
 $B \raw PV$ decays than in $B \raw PP$ decays. This is consistent 
 with theoretical predictions~\cite{sum} that uses factorization
 which predicts destructive (constructive) interference between 
 penguin operators of opposite chirality for $B\to K\rho$ ($B\to K\pi$), 
 leading to a rather small (large) penguin contribution in these decays.

We gratefully acknowledge the effort of the CESR staff in providing us with
excellent luminosity and running conditions.
This work was supported by
the National Science Foundation,
the U.S. Department of Energy,
the Research Corporation,
the Natural Sciences and Engineering Research Council of Canada,
the A.P. Sloan Foundation,
the Swiss National Science Foundation,
the Texas Advanced Research Program,
and the Alexander von Humboldt Stiftung.

\begin{figure} [hhh]
\begin{center}
\includegraphics*[width=10cm,height=10cm]{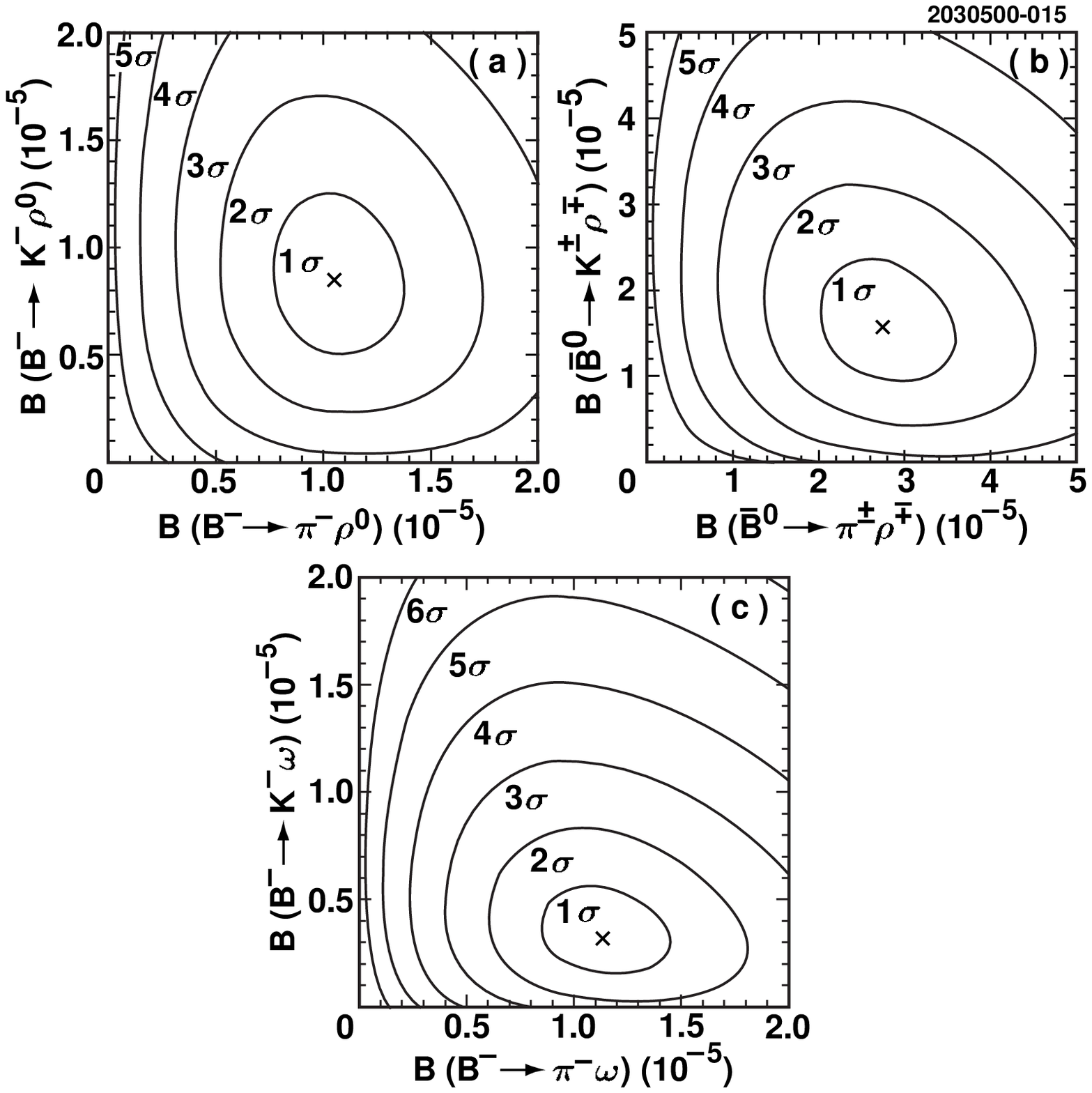}
\end{center}
\label{fig01}
\caption{Likelihood contours at $n$ standard deviations ($\sigma$) 
         of branching fractions for
         $\bm\raw\pim\rhoz$, $\km\rhoz$ (a), 
         $\bzb\raw\pipm\rhomp$, $\kpm\rhomp$ (b) 
         and $\bm\raw\pim\omg$, $\km\omg$ (c).}
\end{figure}


\begin{figure} [htbp]
\begin{center}
\includegraphics*[width=10cm,height=10cm]{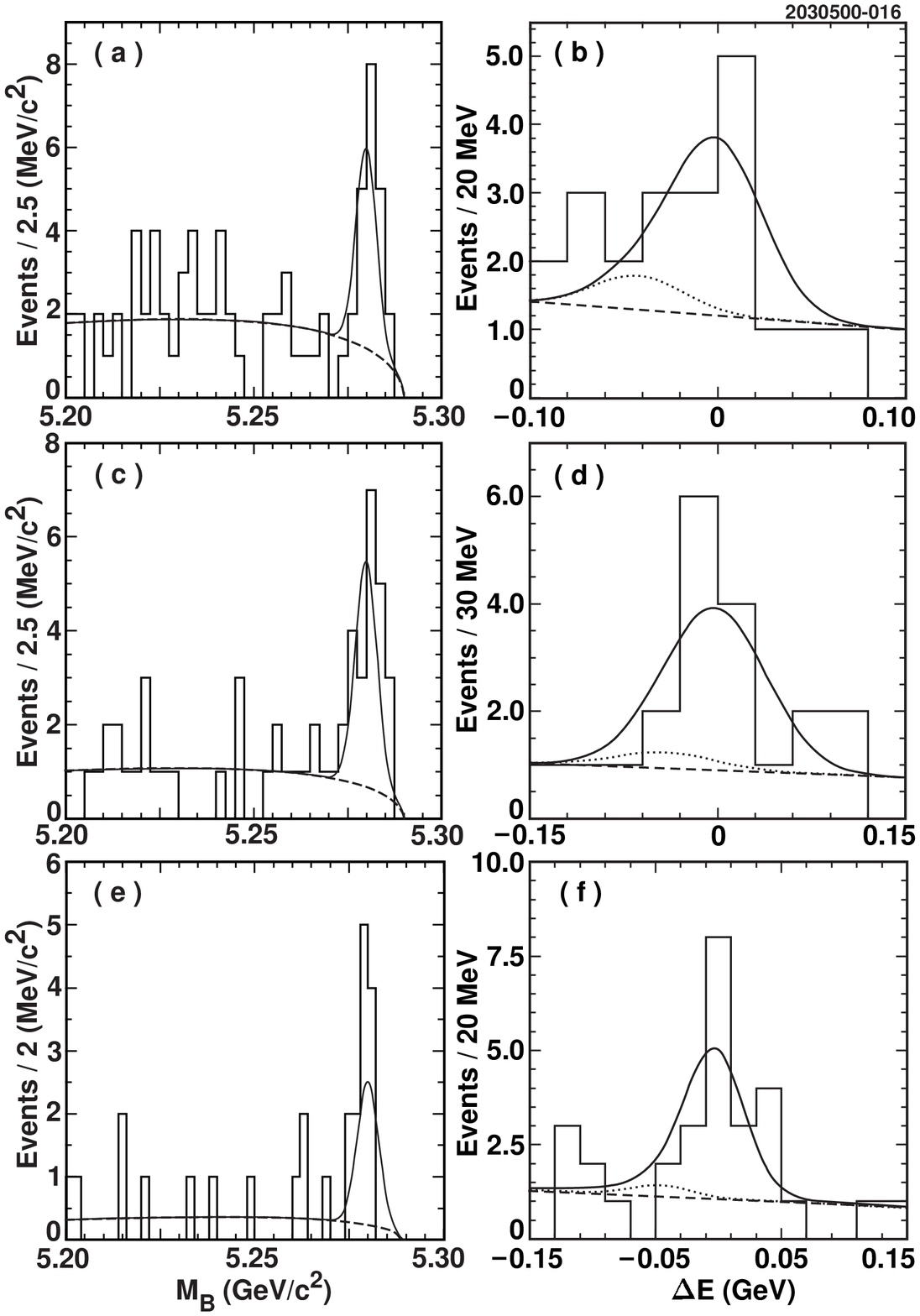}
\end{center}
\label{fig02}
\caption{Projection plots in $M_{B}$ and  $\Delta E$ for
         $\bm\raw\pim\rhoz$ (a,b), $\bzb\raw\pipm\rhomp$ (c,d) 
         and $\bm\raw\pim\omg$ (e,f).
         The histograms show the data while the solid lines represent the
         overall fit to the data scaled to account for the extra 
         requirement on event probability applied to make the projection. 
         The dashed lines represent the continuum and the dotted lines
         on top of the continuum represent the other $B$ components 
         ($\bm\raw\km\rhoz$, $\bzb\raw\kpm\rhomp$ and
         $\bm\raw\km\omg$) in the simultaneous fits.} 
\end{figure}

\end{document}